\documentclass[%
reprint,
superscriptaddress,
amsmath,amssymb,
aps,
pra,
]{revtex4-1}

\usepackage{graphicx}
\usepackage{dcolumn}
\usepackage{bm}
\usepackage{color}

\begin{document}
	
	
	\title{High dimensional measurement device independent quantum key distribution on two dimensional subspaces}
	
	\author{Luca Dellantonio}
    \email{luca.delantonio@nbi.ku.dk}
	\affiliation{The Niels Bohr Institute, University of Copenhagen, Blegdamsvej 17, DK-2100 Copenhagen \O, Denmark}
	\affiliation{Center for Hybrid Quantum Networks (Hy-Q), Niels Bohr Institute, University of Copenhagen, Blegdamsvej 17, DK-2100 Copenhagen \O, Denmark}
	\author{Anders S. S\o rensen}
	\affiliation{The Niels Bohr Institute, University of Copenhagen, Blegdamsvej 17, DK-2100 Copenhagen \O, Denmark}
	\affiliation{Center for Hybrid Quantum Networks (Hy-Q), Niels Bohr Institute, University of Copenhagen, Blegdamsvej 17, DK-2100 Copenhagen \O, Denmark}
	\author{Davide Bacco}
    \email{dabac@fotonik.dtu.dk}
	\affiliation{CoE SPOC, DTU Fotonik, Dep. Photonics Eng., Technical University of Denmark, Orsteds Plads 340, Kgs.~Lyngby, 2800 Denmark}
	\date{\today}
	
	\begin{abstract}
		Quantum key distribution (QKD) provides ultimate cryptographic security based on the laws of quantum mechanics. For point--to--point QKD protocols, the security of the generated key is compromised by detector side channel attacks.
		This problem can be solved with measurement device independent QKD (mdi--QKD). However, mdi--QKD has shown limited performances in terms of the secret key generation rate, due to post--selection in the Bell measurements. We show that high dimensional (Hi--D) encoding (qudits) improves the performance of current mdi--QKD implementations. The scheme is proven to be unconditionally secure even for weak coherent pulses with decoy states, while the secret key rate is derived in the single photon case. Our analysis includes phase errors, imperfect sources and dark counts to mimic real systems. Compared to the standard bidimensional case, we show an improvement in the key generation rate. 
		
	\end{abstract}

\maketitle

\subsection*{Introduction}
Digital security is important for several aspects of modern life. Classical cryptography only promises to make decryption hard, but not impossible. On the contrary, quantum key distribution (QKD) is based on the laws of physics, theoretically allowing parties to share cryptographic keys in an unconditionally secure way \cite{DiQKD2}. However, several physical requirements have to be satisfied to provide unconditional security, and most experimental implementations of QKD have proven to be vulnerable to attacks \cite{Hacking1,Hacking2,Hacking3,Hacking4,Hacking5,Hacking6,Hacking7,Hacking8,Hacking9,Hacking10}. These attacks mainly exploit weaknesses in the detectors, whereas the sources are less vulnerable. To overcome this limitation, device independent (di--QKD) \cite{DiQKD1,DiQKD2,DiQKD3,PirandolaSideChannel}, and measurement device independent QKD (mdi--QKD) \cite{Lo2012} were introduced to decrease the reliance on the physical setup. While di--QKD remains challenging due to technical limitations, including the need for extremely efficient detection \cite{DiQKD2}, mdi--QKD is ready to be implemented in real networks.

Mdi--QKD was introduced by Lo et al. in Ref. \cite{Lo2012}. Here, the two parties Alice and Bob only use photon sources, while the detection is performed by a third party, Charlie. 
Different degrees of freedom have been used to demonstrate the feasibility of this scheme (e.g.: polarization, phase, time, and space)~\cite{Lo2012,bass2001handbook,TFmdiQKD}. Compared to other QKD protocols, however, mdi-QKD has shown low key generation rates.
To reduce this limitation, high-dimensional (Hi--D) encoding can be used to improve the photon information efficiency (PIE) \cite{chau2016qudit}.
Recent results have shown how spatial or temporal modes can be used to increase the dimension of the Hilbert space \cite{Bacco2016,BaccoSpace1,Liu2013,Islame1701491,Englund1} for standard QKD. We propose a protocol, where Alice and Bob generate qudits (quantum states in $N$--dimensions) encoded in different paths or time slots of the photons. These photons then interfere at Charlie's Beam Splitters (BS), as shown in Fig.~\ref{fig:Fig_setup}. As discussed below, the measurement projects the qubits into a two dimensional subspace, which can be used for QKD. In the following, we analyse this high dimensional mdi--QKD protocol, considering the main sources of errors, such as, imperfect photon generation, dark counts and (unknown) phase shifts. We prove that high dimensional mdi--QKD is unconditionally secure for coherent states with the decoy state technique \cite{Lo2012,DecoyState}, and analyse the key generation rate for single photon sources. In analogy to a similar result for standard QKD \cite{Englund1}, we find that our Hi--D mdi--QKD protocol is advantageous, particularly in the detector saturation regime, where the time between photon clicks at Charlie's detectors is comparable to the detectors' dead time $\tau_{d}$. We study the protocol both for time and space encoding, and analyse the practical constraints that make one encoding better than the other. A different Hi--D mdi--QKD scheme was proposed in Ref. \cite{chau2016qudit}, but remains experimentally unfeasible, since discriminating Bell states in high dimensions is impossible by simple means \cite{calsamiglia,Yoran}. In comparison, our protocol can be implemented without significant increase in the complexity of existing setups. In particular, for weak coherent states and time encoding, no change in the hardware is required.

\subsection*{Protocol definition}

\begin{figure}
   \centering
    {\includegraphics[width=8.5cm]{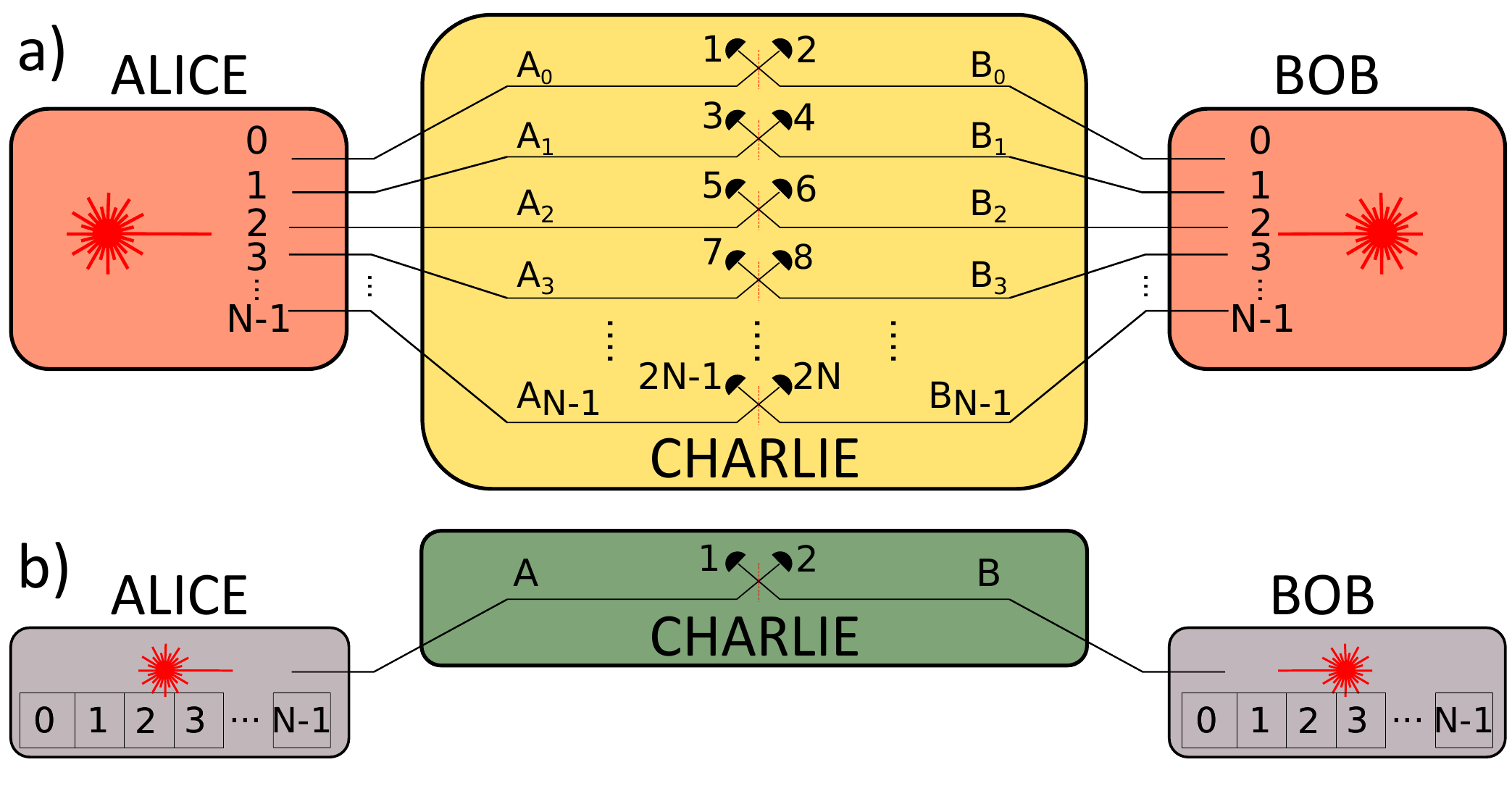} }
    \caption{Schematic of the proposed setup for Hi-D-mdi QKD. {\it (a)} Space is used to encode information in different paths (multi-core fibers can be used as transmission channels). $2N$ single photon detectors are necessary for this configuration. {\it (b)} Time encoding scheme, where different time-slots are used to encode the qudits. The number of detectors is independent of the dimension $N$.}
\label{fig:Fig_setup}
\end{figure}

Most QKD protocols are based on mutually unbiased bases (MUBs). Usually, the computational $Z$ basis ($\left\lbrace \lvert 0 \rangle,\lvert 1 \rangle \right\rbrace$ for qubits) is less susceptible to errors than the $X$ basis ($\left\lbrace \lvert \phi_{0} \rangle ,\lvert \phi_{1} \rangle  \right\rbrace$, with $\lvert \phi_{0} \rangle = (\lvert 0 \rangle+\lvert 1 \rangle)/\sqrt{2}$ and $\lvert \phi_{1} \rangle=(\lvert 0 \rangle-\lvert 1 \rangle)/\sqrt{2}$). This is also the case for the encodings in Fig.~\ref{fig:Fig_setup}, where different wave packets may dephase, but are unlikely to switch from one bin to another. Thus, the $Z$ basis is used for key generation, and the $X$ basis for error estimation. 
Generalizations of the $Z$ and $X$ bases are, respectively, $\left\lbrace \lvert 0 \rangle, \lvert 1 \rangle,...,\lvert N-1 \rangle  \right\rbrace$, and $\left\lbrace \lvert \phi_{0} \rangle,...,\lvert \phi_{N-1} \rangle  \right\rbrace$. Here, $\lvert \phi_{i} \rangle$ are the $N$ orthonormal superpositions of all the elements of the $Z$ basis, with equal and real weights. As an example, for $N=4$:
\begin{subequations}
\begin{align}
\lvert \phi_{0} \rangle = & \frac{1}{2}\left( \lvert 0 \rangle + \lvert 1 \rangle + \lvert 2 \rangle + \lvert 3 \rangle \right),\\
\lvert \phi_{1} \rangle = & \frac{1}{2}\left( \lvert 0 \rangle - \lvert 1 \rangle - \lvert 2 \rangle + \lvert 3 \rangle \right),\\
\lvert \phi_{2} \rangle = & \frac{1}{2}\left( \lvert 0 \rangle + \lvert 1 \rangle - \lvert 2 \rangle - \lvert 3 \rangle \right),\\
\lvert \phi_{3} \rangle = & \frac{1}{2}\left( \lvert 0 \rangle - \lvert 1 \rangle + \lvert 2 \rangle - \lvert 3 \rangle \right).
\end{align}
\end{subequations}

Our $N$ dimensional mdi--QKD protocol for two MUBs is given by the procedure: 
\begin{enumerate}
\item Alice and Bob choose, with probability $P_{b}\in\left( 0,1 \right)$, the $Z$ basis and with probability $1-P_{b}$ the $X$ basis.
\item Alice and Bob randomly generate one of the $N$ qudits in the chosen basis, and send it to Charlie.
\item Whenever Charlie gets a coincidence click of two detectors, he publicly announces the outcome of his measurement. Otherwise, the event is discarded.
\item Steps (1) to (4) are repeated, to have enough statistics to estimate the quantum bit error rate (QBER), and sufficiently many bits of key.
\item Alice and Bob announce their bases, and estimate the QBER. If the QBER is too high, they abort the protocol.
\item Alice and Bob proceed with classical error correction and privacy amplification.
\end{enumerate}

For simplicity (when not otherwise specified), we describe the protocol in the space encoding of Fig.~\ref{fig:Fig_setup}(a), with straightforward generalization to the time encoding. Assume first that Alice and Bob both choose the $Z$ basis. Whenever they send the same element $\lvert i \rangle$, two photons arrive at the same BS and bunch together. There is thus no coincidence event, and the outcome is discarded. When Alice and Bob generate different states $\lvert i \rangle$ and $\lvert j \rangle$ ($i\neq j$), these photons necessarily end up in different detectors, and Charlie gets a coincidence click. The measurement collapses the state onto the two--dimensional space $\lbrace \lvert i\rangle_{A} \otimes \lvert j\rangle_{B};\lvert j\rangle_{A}\otimes \lvert i\rangle_{B} \rbrace$, with the first state being Alice's, the latter Bob's. An eavesdropper Eve cannot distinguish whether Alice sent the state $\lvert i \rangle$ and Bob $\lvert j \rangle$ or vice versa, and thus can only guess with $50\%$ probability the bit of key. In the $X$ basis interference only allows half of all possible coincidence clicks to happen, and this permits determining the QBER relative to all two dimensional subspaces. For example, consider the case $N=2$, and assume that both Alice and Bob send states with the same phase. Then, only coincidences on the same side of the BSs of Fig.~\ref{fig:Fig_setup}(a) are allowed. If Alice and Bob choose different phases, opposite outcomes are permitted. This concept is generalizable to $N>2$, considering that the detection collapses the state onto a two dimensional subspace, so that only the relative phases within this subspace matter. Alice and Bob can thus determine the contributions $\epsilon_{x}^{i,j}$ to the QBER $\epsilon_{x}$, where $i,j=0,...,N-1$ are all possible indices of the $2$--dimensional subspaces of the composite Hilbert space. For finite key length and high dimensions, there may be insufficient statistics to estimate each individual error rate $\epsilon_{x}^{i,j}$. In this case, the QBER can be determined by merging all $X$ measurements into a single error rate $\epsilon_{x}$. The QBER for the $N$ dimensional protocol can thus be estimated with the same resources as for the standard $2$ dimensional protocol \cite{FiniteKey}. If the error rates $\epsilon_{x}^{i,j}$ are different (e.g. due to different detectors), a better key rate can be obtained by treating the errors independently. For simplicity, we restrict ourselves to the simplest strategy and only consider a single error rate $\epsilon_{x}$.

\subsection*{Secret key rate}

We first prove that our Hi--D protocol is unconditionally secure, both for single photon sources and for coherent states with the decoy state method \cite{DecoyState}. Then, we investigate all elements of the setup -- sources, channels and detectors -- to determine the QBER and raw key generation rate per application of the protocol ($R_{p}$) in the single photon case and for realistic experimental conditions. Finally, we consider the detector saturation regime.

In order to prove that Hi--D mdi--QKD is unconditionally secure, we show that the security of the $N$ dimensional protocol follows from the two dimensional case \cite{Lo2012,Inamori2002,Hwang2003,Wang2005}.
The key argument is that, whenever Charlie announces a coincidence click, the wave function is projected onto a two dimensional subspace, with all other states being erased by the measurement. As an example, consider Fig.~\ref{fig:Fig_setup}(a), and assume that one of detectors $1$ and $2$ and one of detectors $7$ and $8$ click. The system is thus projected onto the Bell states $ (\lvert 0 \rangle_{A} \lvert 3 \rangle_{B} \pm \lvert 3 \rangle_{A} \lvert 0 \rangle_{B})/\sqrt{2}$, with the sign determined by the parity of the measurement (clicks in $1$ \& $7$ or $2$ \& $8$ lead to a plus, $1$ \& $8$ or $2$ \& $7$ to a minus). It follows that, if Alice and Bob both chose the $X$ basis, all states other than $\lvert 0 \rangle$ and $\lvert 3 \rangle$ are erased by the measurement. On the other hand, if the $Z$ basis was used, the parties had to have chosen these particular states as qudits. Every successful realization of the Hi--D protocol, is thus equivalent to an application of the two dimensional protocol, with the specific states identified by Charlie's measurement. 

To complete the security proof, we follow Ref. \cite{Lo2012} and consider the virtual qudit approach \cite{gottesman2004}. We imagine that both parties prepare an entangled state of two qudits, of which one is sent to Charlie, and the other (the virtual one) is kept. The travelling photons are then encoded in the basis states by measuring the virtual qudits. Since these measurements can be postponed till after Charlie's outcome is revealed, and since this outcome projects the state onto a two dimensional subsystem, the protocol is equivalent to the entanglement based protocol for qubits \cite{Chau1999,ShorSecurity}. 

The secret key rate $r$ can be derived from Ref. \cite{Lo2012,gottesman2004,lo2005efficient,koashi2005}:

\begin{equation}
r=R\left[ 1-H\left( \epsilon_{x} \right) - f \left( \epsilon_{z} \right)H\left( \epsilon_{z} \right)\right], \label{eq:SecRate2D}
\end{equation}
where $R$ is the raw key rate, $f(x)\geq 1$ is an inefficiency function for the error correction, and $H(x)$ the binary entropy. 
The same security proof can be adapted to the case of weak coherent pulses with the decoy state technique \cite{Lo2012}. Since the measurement collapses the system to a two dimensional subspace, high--dimensional entanglement cannot be fully exploited with the current settings. It is thus not surprising that the Hi--D protocol can be described in terms of standard mdi--QKD protocols. However, as we will see in the following, our protocol still allows for improvements. 

With the protocol proven unconditionally secure, we now estimate the key rate taking into account realistic sources, channels and detectors. Above and in the following we assumed identical channels and detectors \cite{Terza}.

(\textit{Sources}) In Hi--D mdi--QKD both Alice and Bob are required to generate qudits. These Hi--D photons have to interfere to generate the key, and therefore need to be identical. We quantify the errors introduced by distinguishable photons, assuming different shapes of the emitted photons. This can be described by expanding Alice's state $\lvert i_{A} \rangle_{A}$ ($i = 0,...,N-1$) in terms of Bob's wave function according to $\lvert i_{A} \rangle_{A}\rightarrow \beta \lvert i_{B} \rangle_{A} + \sqrt{1-\lvert \beta \rvert^{2}}\lvert I \rangle_{A}$, where $\lvert I \rangle_{A}$ shares the encoding of Bob's state (meaning that is in the same path/time slot), but is in one or more modes other than $\lvert i_{B} \rangle_{A}$. If both parties use the $Z$ basis, there should never be coincidences between detectors associated with the same BS in Fig.~\ref{fig:Fig_setup}(a), and if the photons are in different paths it does not matter if they are distinguishable. Hence, the influence of distinguishable photons can be identified and never leads to errors in the key rate. However, for the $X$ basis, there is a probability $\lvert \beta \rvert^{2}$ that the photons interfere correctly, and a probability $1-\lvert \beta \rvert^{2}$ that they click at random detectors, thus incrementing the QBER $\epsilon_{x}$ by $(1-\lvert \beta \rvert^{2})/2$. 

(\textit{Channels}) The most general errors affecting qudits in transmission lines are bit--flips and phase--shifts \cite{nielsen2000}. We neglect the first ones, since the probability that a photon disappears and reappears in another spatially or temporally separated slot is small \cite{Liu2013,BaccoSpace1}. Instead, within the transmission channel any state $\lvert i \rangle_{J}$ acquires a random phase, such that $\lvert i \rangle_{J} \rightarrow e^{i\theta_{i}^{J}}\lvert i \rangle_{J}$. Here, $i=0,...,N-1$ and $J=A,B$ indicates whether the qudit was generated by Alice or Bob. Like before, the $Z$ basis is unaffected by phase noise, since bits of key are only exchanged when photons do not interfere. However, for any pair of elements in the $X$ basis, interference prevents half of the allowed coincidence clicks. Whenever phase noise affects the qudits, wrong clicks happen with a probability $\left[1-\langle \cos \left( \theta_{i}^{A} - \theta_{j}^{A} - \theta_{i}^{B} + \theta_{j}^{B} \right) \rangle\right]/2$, with $i\neq j$ (the case $i=j$ is automatically discarded). To quantify this effect, a noise model for the random variables $\theta_{i}^{A} - \theta_{j}^{A}$ and $\theta_{j}^{B}  - \theta_{i}^{B}$ is required. Different models are better suited for different transmission lines and encoding schemes. In the space one, we consider a homogeneous situation, such that relative phases $\theta_{i}^{A} - \theta_{j}^{A}$ and $\theta_{j}^{B} - \theta_{i}^{B}$ are Gaussian distributed, with zero average and identical variance $\sigma^{2}$. In the time domain, phase drifts in the sources can be added as independent noise contributions in this model. Here, we assume white noise between subsequent pulses, such that the variances of $\theta_{i}^{A} - \theta_{j}^{A}$ and $\theta_{j}^{B} - \theta_{i}^{B}$ are $\lvert i-j \rvert \sigma^{2}$. Alternatively, if the interferometer is slowly drifting, an appropriate model would be $\lvert i-j \rvert^{2}\sigma^{2}$.

(\textit{Detection}) For long distances, dark counts prevail over real clicks, increasing the QBER. We define $P_{dc}$ the probability that a single detector clicks without a photon, and $P_{s} = \eta 10^{-\alpha_{0} d /10}$ the probability that a photon arrives at a detector and clicks. Here, $\eta$ is the detector's efficiency, $\alpha_{0}$ the fiber loss coefficient and $d$ the distance separating both Alice and Bob from Charlie. In the $Z$ basis, Alice and Bob verify if Charlie's announcement is compatible with the qudit they sent. A wrong bit of key is shared if and only if Alice and Bob send the same state, and a bit--flip (induced by dark counts) occurs. If none or one photon arrives, a random bit of key is shared with probabilities $4\frac{N-1}{N}(1-P_{s})^{2}P_{dc}^{2}(1-P_{dc})^{2N-2}$ (0 photons arrive) and $4\frac{N-1}{N}P_{s}(1-P_{s})P_{dc}(1-P_{dc})^{2N-2}$ (1 photon arrives). In case both photons click at the detectors, the probability to share a correct bit is $\frac{N-1}{N}P_{s}^{2}(1-P_{dc})^{2N-2}$. A wrong bit is produced by two photons bunching together and a different detector firing, which happens with a probability $2\frac{N-1}{N}P_{s}^{2}P_{dc}(1-P_{dc})^{2N-2}$. From these, it is possible to find how many wrong bits of key are shared on average, and thus the QBER $\epsilon_{z}$ and the raw rate per application of the protocol $R_{p}$ in the $Z$ basis. 

\begin{figure}
  \centering
   \includegraphics[width=8cm]{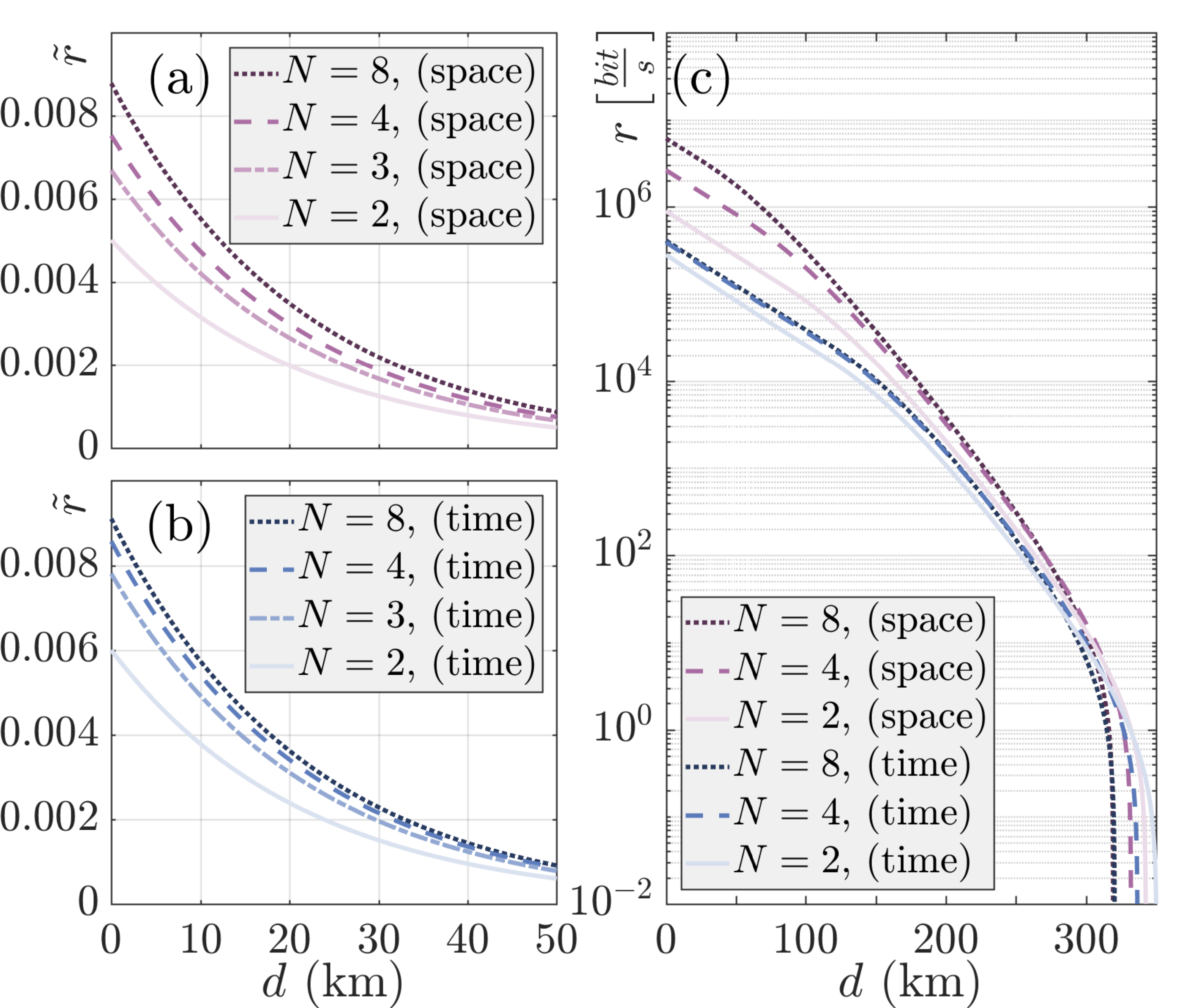}
   \caption{Secure key rate as a function of distance. Plain lines refer to $N=2$, dash--dotted lines to $N=3$, dashed lines to $N=4$ and dotted lines to $N=8$. \textit{(a,b)}: No detector dead time, $\tau_{d}=0 $. The secret key rate without detector dead time $\tilde{r}$, is found using Eq.~\eqref{eq:SecRate2D}, with $R$ substituted by $R_{p}$, i.e.: $\tilde{r}$ is in $bit$ per application of the protocol. \textit{(c)}: Secret key rate per second $r$ as a function of distance. The dead time is $\tau_{d}=20$ n$s$, and the minimum pulse separation $\tilde{T}_{p} = 200$ ps ($\tau_{d}/\tilde{T}_{p}=100$). Common parameters are: $P_{dc}=1\cdot 10^{-6}$, $f(\epsilon_{z})=1$, $\lvert \beta \rvert^{2} = 0.85$, $\eta = 0.145$ and $\sigma$ equal to $0.175$ (time) or $0.325$ (space). $\sigma$ is chosen such that, for $N = 2$ when only including dephasing, there is a QBER $\epsilon_{x}$ of $1.5 \%$ (time) or $5\%$ (space).}
   \label{fig:KeyRate}
\end{figure}

We now explicitly calculate the QBER $\epsilon_{x}$ in the $X$ basis, including phase noise and distinguishability. If no photons arrive at Charlie, half the coincidence clicks are correct, half wrong, both occurring with probability $(1-P_{s})^{2}N(N-1)P_{dc}^{2}(1-P_{dc})^{2N-2}$. With a single photon clicking, the probability to have a correct or wrong coincidence click is $2P_{s}(1-P_{s})(N-1)P_{dc}(1-P_{dc})^{2N-2}$. When both photons click at Charlie's detectors, the probabilities for the outcome to be correct or wrong are $P_{s}^{2}(1-P_{dc})^{2N-2}\left[ P_{\text{good}}^{(X)} + (N-1)P_{dc} P_{\text{double}}^{(X)}\right]$ and $P_{s}^{2}(1-P_{dc})^{2N-2}\left[ P_{\text{bad}}^{(X)} + (N-1)P_{dc} P_{\text{double}}^{(X)}\right]$, respectively. Here, $P_{\text{double}}^{(X)} = (1+\lvert \beta \rvert^{2})/N$ is the probability that both photons end up in the same detector. $P_{\text{bad}}^{(X)} = \left[ N(N-1) - 2\lvert \beta \rvert^{2} f_{N} \right]/(2N^{2})$ and $P_{\text{good}}^{(X)} = \left[ N(N-1) + 2\lvert \beta \rvert^{2} f_{N} \right]/(2N^{2})$ are the probabilities to have or not have the photonic interference spoiled by phase noise and distinguishability. The function $f_{N}$ depends on the considered phase noise model. For the space encoding, we find $f_{N} = N(N-1) e^{-\sigma^{2}}/2$. For the time, $f_{N} = \left[ N\left(1 - e^{-\sigma^{2}} \right) + e^{-N \sigma^{2}} - 1 \right]/\left[ 2 \sinh \left(\sigma^{2}/2\right) \right]^{2}$. With these results, it is possible to find how many bits of key are wrong on average, and thus the QBER $\epsilon_{x}$ in the $X$ basis.

By merging the results above for sources, channel and detection imperfection, we derive Fig.~\ref{fig:KeyRate}(a,b), where the secret key rate per application of the protocol is determined using Eq.~\eqref{eq:SecRate2D}, with $R$ substituted by the raw key rate per application of the protocol $R_{p}$. From the plot we find the advantage of Hi--D mdi--QKD, as compared to standard mdi--QKD. The probability that Alice and Bob send the same state $\lvert i \rangle$ (resulting in a useless event) asymptotically goes to zero. This implies that, for small $P_{dc},$  the performance is improved by a factor $2(N-1)/N$ compared to the standard mdi--QKD protocol, where half of the events are lost even if Alice and Bob select the same basis.



In the following, we study the regime where the detector's dead time $\tau_{d}$ is comparable to the timescale at which photons click at Charlie's detectors, and dark counts are negligible. We assume that during $\tau_{d}$ Alice and Bob send $n$ pulses separated by $T_{p} = \tau_{d}/n$. In this regime, ordinary QKD has proven to gain advantage from high dimensional encoding \cite{Englund1,EnglundEff}. In the following, we extend this result to mdi--QKD, considering space and time encodings separately. 

\begin{figure}
  \centering
   \includegraphics[width=8cm]{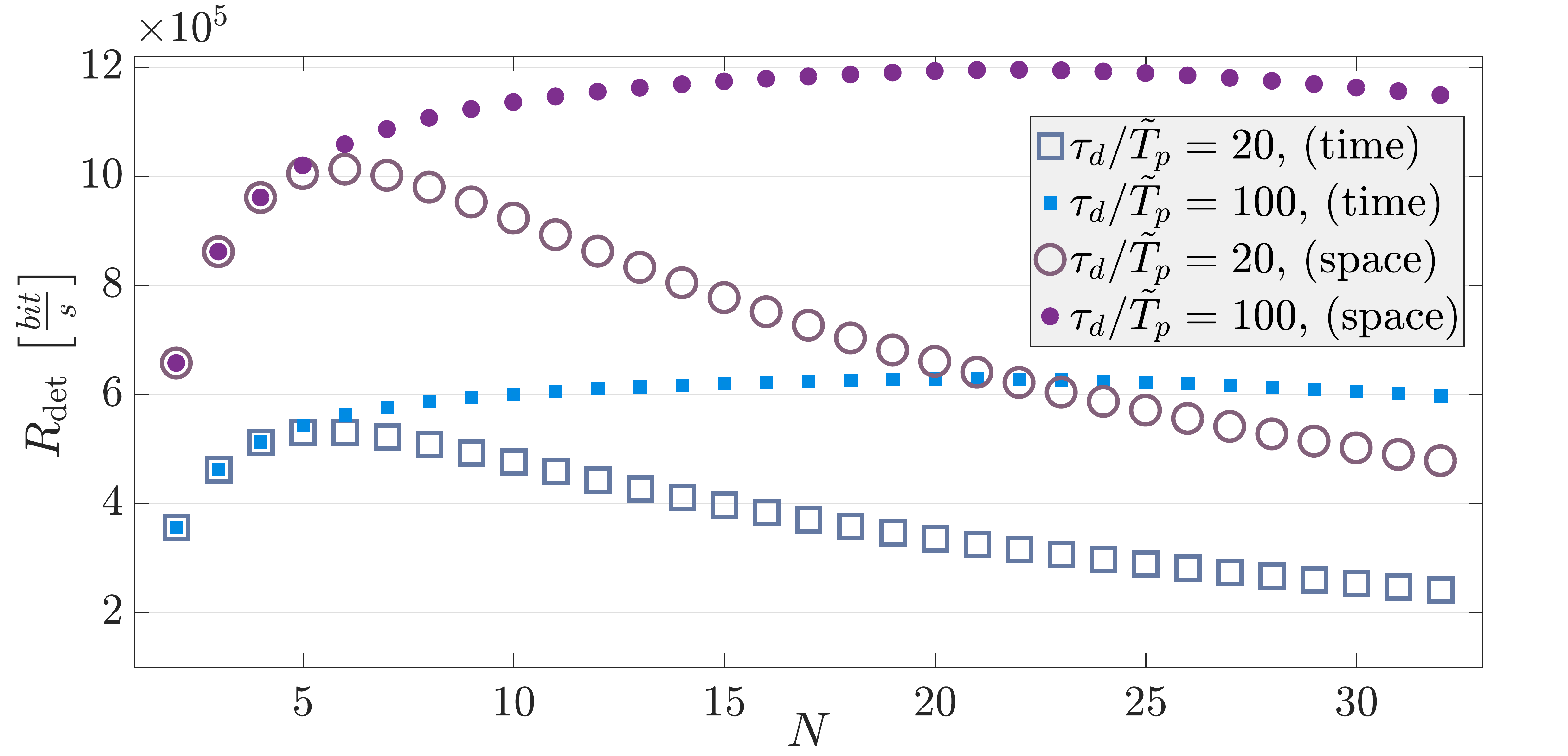}
   \caption{Raw key per detector $R_{\text{det}} = R/ n_{\rm det}$ as a function of the dimension $N$, in the detector saturation regime. Violet circles (full and empty) are used for the space encoding. Blue squares (full and empty) for the time encoding.  The number of pulses $n$ within $\tau_d$  is optimized to achieve the highest rate. The maximum possible number of qubits $\tau_{d}/\tilde{T}_{p}$ is either equal to $20$ (empty circles and squares) or $100$ (full circles and squares). $P_{s}= 0.2$, $\tau_{d} = 20$ n$s$, and $n_{\rm det}= 2N$ (space) or $n_{\rm det}=2$ (time).}
   \label{fig:Saturation}
\end{figure}

(\textit{Space}) For any dimension $N$ of the Hilbert space, $2N$ detectors are used (see Fig.~\ref{fig:Fig_setup}). The probability per pulse $P_{\text{hit}}$ that a detector is hit by a photon is $P_{\text{hit}} = \frac{1}{2N}\left[ 2P_{s}(1-P_{s}) + P_{s}^{2}(2N-1)/N \right]$. In the continuous limit ($t\gg T_{p}$), the cumulative distribution for a detector being hit within a time $t$ is $1-e^{-P_{\text{hit}}t/T_{p}}$. From this, the probability $P_{\text{alive}}$ that a detector is not dark can be found to be $P_{\text{alive}} = P_{\text{hit}}^{-1}/(P_{\text{hit}}^{-1}+n)$, where we assume that a detector remains dark for a time $\tau_{d}$, no matter how many photons arrive while it is dark. The average number of raw bits $N_{\text{raw}}$ exchanged during a dead time $\tau_{d}$ is therefore
\begin{equation}\label{eq:NbitsSat}
N_{\text{raw}} = \frac{\tau_{d}}{T_{p}}\frac{(N-1) P_{s}^{2}P_{\text{alive}}^{2}}{N}.
\end{equation}
Maximizing $N_{\text{raw}}$ with respect to $T_{p}$, we find the maximum of $N_{\text{raw}}$ (assuming $P_{s}\ll N$):
\begin{equation}\label{eq:NbitsSpace}
\frac{N_{\text{raw}}^{(M)}}{\tau_{d}}=\max\limits_{T_{p}}\left\lbrace \frac{N_{\text{raw}}(T_{p},P_{s},N)}{\tau_{d}} \right\rbrace = \frac{P_{s}(N-1)}{4 \tau_{d}}.
\end{equation}
(\textit{Time}) In the time encoding, two detectors are used [see Fig.~\ref{fig:Fig_setup}(b)], and the minimum time separation between two consecutive qudits is $N T_{p}$. Following the same procedure outlined above, we find $N_{\rm raw}$, that is the same as in Eq.~\eqref{eq:NbitsSat}, but divided by a factor $2$. This follows from the fact that during a train of $N$ pulses, the same detector cannot click twice, leading to a better performance of the space protocol for short distances (see Fig.~\ref{fig:Saturation}). The maximum number of bits exchanged during the detector's dead time $\tau_{d}$ is thus ($P_{s} \ll N$):
\begin{equation}\label{eq:NbitsTime}
\frac{N_{\text{raw}}^{(M)}}{\tau_{d}}= \frac{P_{s}(N-1)}{8\tau_{d} N}.
\end{equation}

Including the results found for the saturation regime, and limiting the interval $T_{p}$ between consecutive pulses to some minimal value $\tilde{T}_{p}$, the raw key rate $R$ can finally be determined to be: 
\begin{equation}\label{eq:SecRate}
R = \frac{N_{\text{raw}}^{(M)}}{\tau_{d}}R_{p}
\end{equation}
where the raw key rate per application of the protocol $R_{p}$ assumes no detector dead time $\tau_{d} = 0$. Here, $N_{\text{raw}}^{(M)}$ is either Eq.~\eqref{eq:NbitsSpace} (space encoding) or Eq.~\eqref{eq:NbitsTime} (time encoding) when the optimal $T_{p}$ is bigger than $\tilde{T}_{p}$. Otherwise, $N_{\text{raw}}^{(M)}$ is given by Eq.~\eqref{eq:NbitsSat} with the substitution $T_{p}\rightarrow \tilde{T}_{p}$. Since the number of pulses is varied to reach the optimal performance, we evaluate the raw key rate in units of the detector dead time $\tau_{d}$. Therefore, while $R_{p}$ is in $bit/{\rm pulse}$, $R$ is in unit of $bit/s$. 

Since detectors are usually the limiting resource, we renormalize the raw key rate $R$ in Eq.~\eqref{eq:SecRate} with respect to the number of detectors $n_{\rm det}$ employed. This renormalization takes into account that $2N$ detectors could be used to perform $N$ parallel applications of a two dimensional protocol, possibly outperforming the Hi--D setup. The rates per resource are shown in Fig.~\ref{fig:Saturation}, with the plain dots referring to $\tilde{T}_{p}=\tau_{d}/100$, the empty ones to $\tilde{T}_{p}=\tau_{d}/20$. Figure \ref{fig:Saturation} shows that with a limited rate of pulse generation (and thus finite $\tilde{T}_{p}$), there exists an optimal dimension $N_{\rm opt}$ for the best key rate: $N_{\rm opt} = 2 + P_{s} \tau_{d}/\tilde{T}_{p}$  (for $P_{s}\ll N$). For $P_{s} \tau_{d}/T_{p}\gtrsim 1$, we see that with Hi--D mdi--QKD we increase the key rate \textit{per detector}, 
due to the factor $2(N-1)/N$ found above. 

Our work allows, for given experimental conditions, to evaluate \textit{a priori} which is the best setting to be employed in order to achieve the highest secret key rate. As an example, Fig.~\ref{fig:KeyRate}(c) shows the secret key rate $r$ as a function of distance. For these curves, we used Eqs.~\eqref{eq:SecRate} and \eqref{eq:SecRate2D} to determine the raw ($R$) and the secret ($r$) key rates, respectively. With the chosen parameters, for short distances it is better to use Hi--D mdi--QKD in the space encoding, while for very long distances low dimensional time encoding is preferable. Three regimes are visible in the plot: in the central region the rate scales as $P_{s}^{2}$, as two clicks are required. In the detector saturation regime, the probability for the detectors not to be dark is $P_{s}^{-1}$, meaning that the rate is linear in $P_{s}$. Finally, for large distances dark counts prevail, making QKD impossible. Note that for an accurate cost analysis, the number of detectors employed must also be considered, as in Fig.~\ref{fig:Saturation}.

\subsection*{Conclusion}
In conclusion, we have generalized the standard mdi--QKD protocol to higher dimensions $N$. In our analysis we consider the main sources of errors, and we prove the advantages of Hi--D mdi--QKD, particularly in the detector saturation regime. This result improves previous mdi--QKD schemes, allowing for higher communication rates. 
The considered generalization to Hi--D mdi--QKD is only one out of many possibilities. An attractive feature of our proposal, is that it can directly be implemented with existing technology. The protocol works by projecting the state onto a two dimensional Hilbert space, through the Bell state measurement performed by Charlie. Genuine Hi--D Bell state analysers would allow higher key rates, by increasing the PIE and reducing the informations acquired by Eve. However, discriminating Bell states with linear optics is challenging, leaving the Hi--D ones inaccessible \cite{calsamiglia}. The proposals in Refs. \cite{smith2018approaching,goyal2014qudit,duvsek2001discrimination} for Hi--D Bell state analysis, may allow for genuine exploitation of high dimensional Bell states, but remain experimentally challenging. The present approach is thus the most attractive from a practical perspective.

\subsection*{Acknowledgements}
We thank D. Pastorello and S. Paesani for fruitful discussions. This work was supported by the Danish National Research Foundation through the Centers of Excellence Hy--Q (ref DNRF139) and SPOC (Silicon Photonics for Optical Communications, ref DNRF123), by the European Union Seventh Framework Programme through the ERC Grant QIOS, and through the People Programme (Marie Curie Actions) under REA grant agreement n$^\circ$ $609405$ (COFUNDPostdocDTU), and by the Danish Council for Independent Research (DFF).


%


\begin{thebibliography}{41}
\bibitem [{Ter()}]{Terza}%
	\BibitemOpen
	\href@noop {} {}\bibinfo {note} {The extension to the asymmetric case in
		which detectors and/or channels are different can be done, by keeping track
		of the different contributions from all possible two dimensional mdi--QKD
		protocols. Eq.~\eqref{eq:SecRate2D} is then generalized by summing over all
		contributions to the raw key rate and the QBER.}\BibitemShut {Stop}%
\end{thebibliography}
\end{document}